\documentclass[twocolumn,amsmath,amssymb,osajnl]{revtex4}
\usepackage{epsfig}

\begin{document}

\title{Improved large-mode area endlessly single-mode photonic crystal fibers}

\author{N.~A. Mortensen,$^*$ M.~D. Nielsen,$^{*\dagger}$ J.~R. Folkenberg,$^*$ A. Petersson,$^*$ and H.~R. Simonsen$^*$}
\affiliation{$^*$Crystal Fibre A/S, Blokken 84, DK-3460 Birker\o d, Denmark\\
$^{\dagger}$Research Center COM, Technical University of Denmark, DK-2800 Kongens Lyngby, Denmark}
%\email{nam@crystal-fibre.com}
% \homepage{http://www.crystal-fibre.com}

%\date{\today}

\begin{abstract}
We numerically study the possibilities for improved large-mode area endlessly single mode photonic crystal fibers for use in high-power delivery applications. By carefully choosing the optimal hole diameter we find that a triangular core formed by three missing neighboring air holes considerably improves the mode area and loss properties compared to the case with a core formed by one missing air hole. In a realized fiber we demonstrate an enhancement of the mode area by $\sim 30\,\%$ without a corresponding increase in the attenuation.
\end{abstract}


\maketitle

Applications requiring high-power delivery call for single-mode large-mode area (LMA) optical fibers. While standard-fiber technology has difficulties in meeting these requirements the new class\cite{knight1996} of all-silica photonic crystal fibers (PCF) has a big potential due to their endlessly single-mode properties \cite{birks1997} combined with (in principle) unlimited large effective areas.\cite{knight1998el} For recent reviews we refer to Refs.~\onlinecite{birks2001,knight2002}.

\begin{figure}[b!]
\begin{center}
\epsfig{file=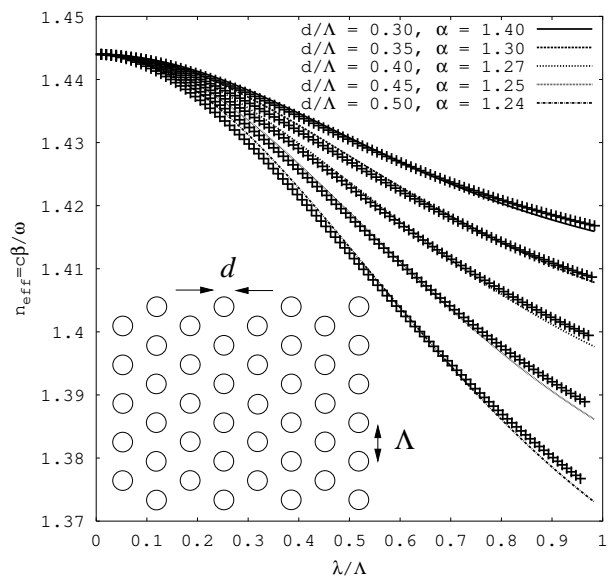, width=0.45\textwidth,clip}
\end{center}
\caption{Effective cladding index for the triangular cladding structure with different hole diameters. The points show numerical results from a fully-vectorial calculation and the solid lines are fits to Eq.~(\ref{fit}) with $\alpha$ as fitting parameter.}
\label{cladding}
\end{figure}

The cladding structure of these PCFs consists of a triangular array of air holes of diameter $d$ and pitch $\Lambda$ corresponding to an air-filling fraction $f=\pi/(2\sqrt{3})(d/\Lambda)^2$. The presence of the air holes results in a strongly wavelength dependent effective index $n_{\rm eff}$ of the cladding and in the short and long wavelength limits we have

\begin{equation}\label{limits}
\lim_{\lambda \ll \Lambda}n_{\rm eff}=n_{\rm si}\;,\; \lim_{\lambda \gg \Lambda}n_{\rm eff}=f\times n_{\rm air}+ (1-f)\times n_{\rm si}\equiv \bar{n}.
\end{equation}
The numerical results in the intermediate regime can be reasonably fitted by {\it e.g.}

\begin{equation}\label{fit}
n_{\rm eff}\approx \bar{n}+(n_{\rm si}-\bar{n})\cosh^{-2}(\alpha \lambda/\Lambda)
\end{equation}
with $\alpha$ of order unity and only weakly dependent on $d/\Lambda$, see Fig.~\ref{cladding}. It is these unusual dispersion properties of the cladding which facilitate design of large-mode area endlessly single-mode optical fibers.\cite{birks1997,knight1998el}

\begin{figure}[t!]
\begin{center}
\epsfig{file=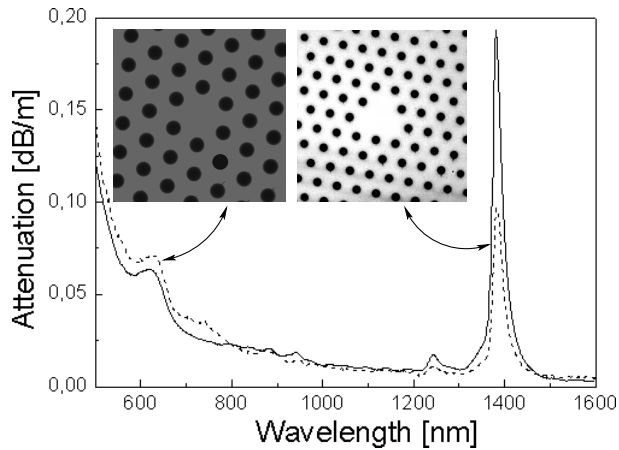, width=0.45\textwidth,clip}
\end{center}
\caption{Attenuation for a single-rod PCF (left insert) and three-rod PCF (right insert) fabricated under comparable conditions and both measured with a bend-radius of $16\,{\rm cm}$. The single-rod PCF has $\Lambda \simeq 10\,{\rm \mu m}$, $d/\Lambda\simeq 0.45$, and a mode-field diameter around $10.5\,{\rm\mu m}$ whereas the three-rod PCF has $\Lambda \simeq 6\,{\rm \mu m}$, $d/\Lambda\simeq 0.25$, and a mode-field diameter around $12\,{\rm\mu m}$. Though the mode area of the three-rod PCF is enhanced by $\sim 30\,\%$ compared to the single-rod PCF the two types of PCFs have very similar attenuation.}
\label{attenuation}
\end{figure}

In order to confine the light to a core region of high index a defect in the triangular air-hole array is introduced. Normally this is done by leaving out one of the air holes. In the stack-and-pull approach \cite{knight1996} one of the capillaries is replaced by a silica rod, see left insert of Fig.~\ref{attenuation}. By choice the index of the defect can be raised by various doping and depressed-index core has also been studied recently.\cite{mangan2001} 

The single-rod PCF can in principle be kept endlessly single-mode no matter how large a core diameter.\cite{knight1998el} However, when scaling-up the fibre-structure the mode area is increased at the cost of an increased susceptibility to longitudinal modulations \cite{mortensen_ptl} such as {\it e.g.} micro-bending \cite{nielsen2002} and macro-bending \cite{sorensen2001} induced scattering loss. The reason is that in order to increase the mode area the pitch $\Lambda$ is scaled to a large value, but this also implies that $\lambda/\Lambda \ll 1$ and in this limit the core index approaches the cladding index, see Eq.~(\ref{limits}).  Fig.~\ref{cladding} suggests that the decreasing index step may be compensated by increasing the air hole diameter, which can be done up to $d/\Lambda\sim 0.45$ which is the upper limit for endlessly single-mode operation. For a discussion of the particular mumber see {\it e.g.} Refs.~\onlinecite{broeng1999,mortensen2002a,kuhlmey2002}. For LMA PCFs working in the UV and visible regimes this sets an upper limit on the mode areas that can be realized with a reasonable loss and many applications call for an improved LMA PCF design.

The inclusion of more than a single solid rod in the stacking has been used to form multiple-core \cite{mangan2000} and highly birefringent PCFs.\cite{hansen2001} In this work we demonstrate how inclusion of more neighboring solid rods can be used for improved LMA endlessly single-mode PCFs. Intuitively this may not seem to be a promising direction since a reduced value of $d/\Lambda$ is needed to keep the PCF endlessly single-mode. For the birefringent case with two neighboring rods\cite{hansen2001} the limit is $d/\Lambda \sim 0.30$ and for a triangular core formed by three neighboring rods (see right insert of Fig.~\ref{attenuation}) we have found $d/\Lambda \sim 0.25$ as the upper limit for endlessly single-mode operation. However, for a given desired mode area this decrease in $d/\Lambda$ is compensated for by a corresponding smaller value of $\Lambda$. In fact, the edge-to-edge separation $\Lambda-d$ of the holes turns out to be the important length scale rather than the pitch $\Lambda$ itself.

In introducing a multiple number of rods an important question about possible birefringence arises. The structure with a single rod has a six-fold symmetry and though group theory clearly excludes any intrinsic birefringence \cite{white2001} there has been quite some debate based on numerical studies, see {\it e.g.} Ref.~\onlinecite{koshiba2001} and references therein. More generally, group theory predicts that for $m$-fold rotational symmetry and $m>2$ a mode with a preferred direction is one of a pair, see Ref.~\onlinecite{white2001} and references therein. PCFs with a triangular core formed by three neighboring rods have a $3$-fold symmetry and thus no intrinsic birefringence. The non-birefringent property is also confirmed numerically using a fully-vectorial plane-wave method \cite{johnson2001} and any small numerical birefringence originates from a numerical grid with symmetry different from the dielectric structure being studied.

In order to compare the single-rod and three-rod PCFs we study two quantities; {\it i)} the mode-field diameter $\rm MFD$ and {\it ii)} the coupling length $\zeta$ to the cladding. We relate the $\rm MFD$ to the effective area\cite{mortensen2002a}

\begin{equation}\label{Aeff}
A_{\rm eff}= \Big[\int d{\boldsymbol r}_\perp I({\boldsymbol r}_\perp)\Big]^2\Big[\int d{\boldsymbol r}_\perp I^2({\boldsymbol r}_\perp)\Big]^{-1},
\end{equation}
by $A_{\rm eff}=\pi ({\rm MFD}/2)^2$. Here, $I({\boldsymbol r}_\perp)$ is the transverse intensity distribution of the fundamental mode. For a Gaussian mode of width $w$ Eq.~(\ref{Aeff}) gives ${\rm MFD}=2w$ and the intensity distribution in the types of PCF studied in this work can be considered close to Gaussian\cite{mortensen2002a,mortensen2002b} as we also confirm experimentally.

The coupling length (beat length)

\begin{equation}\label{zc}
\zeta=2\pi/(\beta-\beta_{\rm cl})
\end{equation}
between the fundamental mode and the cladding (radiation field) can be used in formulating a low-loss criterion.\cite{love} The additional competing length scales consist of the wavelength and the length scale $L_n$ (or as set $\{L_n\}$ of length scales) for nonuniformity along the fiber and loss will be significant when
\begin{equation}\label{high-loss}
\lambda  \lesssim L_n \lesssim \zeta
\end{equation}
and otherwise loss can be expected to be small. Thus, the shorter a coupling length the lower susceptibility to longitudinal modulations. We emphasize that this criterion does not quantify loss, but it gives a correct parametric dependence of loss for various loss mechanisms. For PCFs the relevance of this criteria was recently confirmed experimentally in the case of macro-bending \cite{mortensen_ptl} and micro-bending \cite{nielsen2002} induced nonuniformities and also in a study of PCFs with structural long-period gratings.\cite{kakarantzas2002}

\begin{figure}[t!]
\begin{center}
\epsfig{file=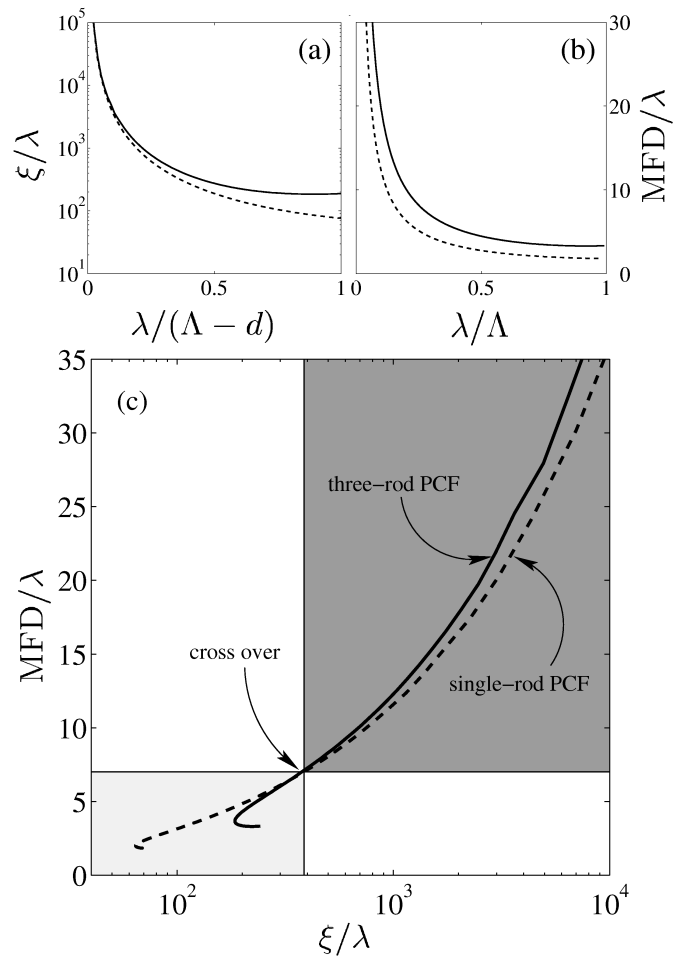, width=0.5\textwidth,clip}
\end{center}
\caption{Comparison of the single-rod (dashed lines) and three-rod (solid lines) PCFs with $d/\Lambda = 0.45$ and $0.25$, respectively. Panel (a) shows the coupling length versus wavelength and panel (b) shows the mode-field diameter as a function of wavelength. In panel (c) the results of panels (a) and (b) are combined to a plot of mode-field diameter versus coupling length.}
\label{numerics}
\end{figure}

In Fig.~\ref{numerics} we compare the single-rod  and three-rod PCFs with $d/\Lambda = 0.45$ and $0.25$, respectively. All numerical results are based on a fully-vectorial solution of Maxwell's equations in a plane-wave basis\cite{johnson2001} and for silica we have for simplicity used $n_{\rm si}=1.444$. Panel (a) shows the coupling length versus wavelength. The normalization by the edge-to-edge separation $\Lambda-d$ of the air holes makes the two curves coincide at short wavelengths ($\lambda \ll \Lambda-d$) which clearly demonstrates that $\Lambda-d$ is the length scale of the fiber structure which determines the susceptibility to longitudinal modulations. Panel (b) shows the mode-field diameter as a function of wavelength and as seen the three-rod PCF provides a larger $\rm MFD$ compared to the single-rod PCF for fixed $\lambda/\Lambda$. Panel (c) combines the results of panels (a) and (b) in a plot of mode-field diameter versus coupling length. At ${\rm MFD}\sim 7\times \lambda$ there is a clear cross over and for ${\rm MFD}\gg  \lambda$ the three-rod PCF is thus seen to be less susceptible to longitudinal modulations compared to the single-rod PCF.

Fig.~\ref{attenuation} shows experimental results for the attenuation of both a single-rod PCF and a three-rod PCF with hole diameters ($d/\Lambda\simeq 0.45$ and $0.25$, respectively) close to the endlessly single-mode limits. The pitches are $\Lambda \simeq 10\,{\rm \mu m}$ and $\Lambda \simeq 6\,{\rm \mu m}$, respectively, so that core sizes are approximately the same. The two PCFs were fabricated by aid of the stack-and-pull method under comparable conditions and both PCFs were found to be endlessly single-mode in a wavelength range of at least $400\,{\rm nm}$ to $1600\,{\rm nm}$. As seen the two PCFs have similar spectral attenuation even though the mode area of the three-rod PCF is enhanced by $\sim 30\,\%$ compared to the single-rod PCF. This demonstrate the improvement by the three-rod PCF.

In conclusion we have found that a triangular core formed by three missing neighboring air holes considerably improves the mode area and/or loss properties compared to the case with a core formed by one missing air hole. This new improved large-mode area endlessly single-mode PCF is important for high-power delivery applications and in a realized fiber we have been able to demonstrate an enhancement of the mode area by $\sim 30\,\%$ without a corresponding change in the loss level.

We acknowledge A. Bjarklev (Research Center COM, Technical University of Denmark) and J. Broeng (Crystal Fibre A/S) for useful discussions. M.~D.~N. is financially supported by the Danish Academy of Technical Sciences.

\end{document}